\documentclass[aip,preprint,amsmath,groupedaddress,superscriptaddress]{revtex4-1}

\usepackage{graphicx}
\usepackage{epstopdf}
\usepackage[hidelinks]{hyperref}
\hypersetup{colorlinks, citecolor=black, linkcolor=black, urlcolor=black}
\usepackage[labelsep=space,justification=RaggedRight]{caption}

\begin{document}

\title{Interfacial Coupling and Electronic Structure of Two-Dimensional Silicon Grown on the Ag(111) Surface at High Temperature}

\author{Jiagui Feng} \affiliation{$^{1}$ Department of Physics and Astronomy, Michigan State University, East Lansing, Michigan 48824-2320, USA}
\author{Sean R. Wagner} \affiliation{$^{1}$ Department of Physics and Astronomy, Michigan State University, East Lansing, Michigan 48824-2320, USA} 
\author{Pengpeng Zhang} \email{zhang@pa.msu.edu} \affiliation{$^{1}$ Department of Physics and Astronomy, Michigan State University, East Lansing, Michigan 48824-2320, USA}

\begin{abstract}

Freestanding silicene, a monolayer of Si arranged in a honeycomb structure, has been predicted to give rise to massless Dirac fermions, akin to graphene. However, Si structures grown on a supporting substrate can show properties that strongly deviate from the freestanding case. Here, combining scanning tunneling microscopy/spectroscopy and differential conductance mapping, we show that the electrical properties of the  ($\sqrt{3}\times\sqrt{3}$) phase of few-layer Si grown on Ag(111) strongly depend on film thickness, where the electron phase coherence length decreases and the free-electron-like surface state gradually diminishes when approaching the interface. These features are presumably attributable to the inelastic inter-band electron-electron scattering originating from the overlap between the surface state, interface state and the bulk state of the substrate. We further demonstrate that the intrinsic electronic structure of the as grown ($\sqrt{3}\times\sqrt{3}$) phase is identical to that of the ($\sqrt{3}\times\sqrt{3}$)R$30^{\circ}$ reconstructed Ag on Si(111), both of which exhibit the parabolic energy-momentum dispersion relation with comparable electron effective masses. These findings highlight the essential role of interfacial coupling on the properties of two-dimensional Si structures grown on supporting substrates, which should be thoroughly scrutinized in pursuit of silicene.

\end{abstract}

\maketitle

\noindent Silicene, a monolayer of Si arranged in a honeycomb structure, has attracted tremendous attention in the past few years as an alternative Dirac system to graphene$^{1-4}$, with the advantages of being readily adapted to the current mainstream Si-based electronics and possessing a strong spin-orbit coupling which may lead to potential applications in spintronics$^{5-7}$. Thus far, freestanding silicene has not been synthesized. Most of the silicene structures were grown on Ag surfaces$^{8-18}$, with a few successes on Zr$\textrm{B}_{2}$$^{19}$ and Ir$^{20}$. Yet, interaction with the substrate may render it rather difficult to probe the intrinsic properties of silicene$^{21-24}$. Furthermore, if the inversion symmetry of the silicene lattice is broken by orbital hybridization with the substrate, it will lead to the breakdown of the Dirac fermion characteristics that are predicted in the freestanding silicene$^{21-24}$.

Among the various superstructures observed on the Ag(111) surface, silicene of the ($\sqrt{3}\times\sqrt{3}$) phase has been claimed to be weakly bound to the substrate and thus maintains the massless Dirac fermions in early reports$^{25-28}$. However, more recent studies found that the ($\sqrt{3}\times\sqrt{3}$) surface reconstruction occurs on multilayered Si instead of a monolayer$^{15-18,29-30}$. In addition, such a ($\sqrt{3}\times\sqrt{3}$) phase resembles both geometrically and electronically the reconstructed Ag on Si(111), suggesting the formation of a surface alloy$^{31-33}$. Two central questions then arise, i.e., what is the bonding configuration in this multilayered structure and how stable is the film. Since it is well known that Si tends to form $sp^3$ hybridization over $sp^2$ at room temperature and atmospheric pressure, one would expect a bulk-like Si structure to form spontaneously as the interaction strength with the substrate surface decays with increasing layer thickness. Indeed, this structure transition was observed recently by low energy electron microscopy and Raman measurements$^{31,34}$.

In this letter, we report the influence of interfacial coupling on the electronic structure and electrical properties of multilayered Si grown on the Ag(111) surface via scanning tunneling microscopy (STM) and spectroscopy (STS) measurements. We observe the growth of a ($\sqrt{3}\times\sqrt{3}$) phase on top of the ($\sqrt{7}\times\sqrt{7}$) interfacial layer, which results in a ($\sqrt{21}\times\sqrt{21}$) Moir\'{e} pattern. We further show that the  intrinsic electronic structure of the ($\sqrt{3}\times\sqrt{3}$) reconstructed surface is identical to that of the Si(111)-Ag($\sqrt{3}\times\sqrt{3}$)R$30^{\circ}$, and both exhibit the parabolic dispersion relation with comparable electron effective masses. However, in few-layer ($\sqrt{3}\times\sqrt{3}$) Si structures, the electron phase coherence length decreases and the free-electron-like surface state gradually diminishes when approaching the interface, suggesting a strong substrate influence on the electrical properties of thin films. We attribute this finding to the inelastic inter-band electron-electron scattering originating from the overlap between the surface state, interface state and the bulk state of the substrate.

\section*{Results and Discussion}

\noindent The growth of Si on the Ag(111) substrate shows a rich phase diagram, among which the ($\sqrt{3}\times\sqrt{3}$) phase is often observed in multilayers grown at a relatively high temperature. As shown in Fig.~S1(a), at $\textrm{320}^\circ$C a nearly complete monolayer of the ($\sqrt{7}\times\sqrt{7}$) superstructure initially forms, consistent with the previous reports$^{9,31,34}$. It is worth noting that ($\sqrt{7}\times\sqrt{7}$) is not a highly ordered phase, rendering assignment of the atomic structure challenging$^{10,35,36}$. Recently, a universal model has been proposed to explain the variety of ($\sqrt{7}\times\sqrt{7}$) structures that have been reported thus far$^{14}$. Essentially, ($\sqrt{7}\times\sqrt{7}$) stands for a superstructure formed between a Si monolayer in the honeycomb lattice (Si($\textrm{1}\times\textrm{1}$)) and the Ag(111) substrate surface  with the disorder driven by strain relaxation. Upon further deposition, the ($\sqrt{3}\times\sqrt{3}$) phase emerges. Fig.~S1(b) illustrates a surface with the co-existence of both structures.

To investigate the evolution of the ($\sqrt{3}\times\sqrt{3}$) reconstructed film and its interplay with both the ($\sqrt{7}\times\sqrt{7}$) Si superstructure and the Ag(111) substrate, we perform STM and STS measurements. Fig.~1(a) shows the observation of the first ($\sqrt{3}\times\sqrt{3}$) atomic layer. The area outlined by the black dotted line is a continuous ($\sqrt{3}\times\sqrt{3}$) film, presumably grown around a defect feature on the Ag(111) substrate. The continuity of the film is illustrated in a zoomed-in image of the area as shown in Fig.~S2. The apparent height of this layer, as measured at the sample bias of $+1.5\ \textrm{V}$ from the Ag substrate surface (the  lowest terrace) to the as grown Si film (the highest terrace) along the green curve in Fig.~1(c), is about $\textrm{1.50}\pm\textrm{0.10} \ \textrm{\AA}$. To exclude the influence of the substrate defect on the apparent height measurement, Fig. S3(a) shows another area of the film grown on a large and clean Ag(111) terrace. As depicted in the line profile, the apparent height of the first ($\sqrt{3}\times\sqrt{3}$) atomic layer is $1.60\ \textrm{\AA}$, within the same range as that obtained in Fig.~1(c). Note that the height measurement depends on the bias applied between the tip and sample due to the different density of states present on the surfaces of the Ag(111) substrate and the ($\sqrt{3}\times\sqrt{3}$) reconstructed adlayer$^{28}$. Fig.~1(b) shows the multilayered ($\sqrt{3}\times\sqrt{3}$) structures with further deposition, where the line profiles along the black, red, and blue traces indicate that the inter-layer spacing of the ($\sqrt{3}\times\sqrt{3}$) reconstructed films beyond the first atomic layer is $\textrm{3.14}\pm\textrm{0.03} \ \textrm{\AA}$, consistent with the d-spacing of bulk Si(111) within the experimental error. We attribute this observation to the formation of $sp^3$ hybridization in the bulk-like multilayered films. The step height of Ag(111) is $\textrm{2.36}\pm\textrm{0.03} \ \textrm{\AA}$, as marked by the black arrow in Fig. 1(c).

High-resolution STM imaging is further performed on the ($\sqrt{3}\times\sqrt{3}$) structures, as depicted by the square boxes in Figs.~1(a) and (b), with well-defined layer numbers. Fig.~2 shows the corresponding first layer ((a)-(c)), second layer ((d)-(f)), third layer ((g)-(i)), and fourth layer ((j)-(l)) topographies at three different sample biases ($+1.5\ \textrm{V}$, $+0.5\ \textrm{V}$, and $-1.0\ \textrm{V}$). The primitive unit cell and surface superstructures are labeled by the red and black diamonds, respectively, with the corresponding diffraction spots illustrated by the red and black circles in the inserted fast Fourier transform (FFT) images. As one can see, on the first and second atomic layers, a pronounced ($\sqrt{21}\times\sqrt{21}$) periodicity in addition to the ($\sqrt{3}\times\sqrt{3}$) pattern is observed in both the STM and FFT images taken at the sample biases of $+1.5\ \textrm{V}$ and $+0.5\ \text{V}$. However, the ($\sqrt{21}\times\sqrt{21}$) superstructure decays with increasing layer thickness, as evidenced by the images taken on the third and fourth atomic layers at the same biases, and eventually disappears on thicker films (see the STM images in Fig.~S4 which were taken on the sixth ($\sqrt{3}\times\sqrt{3}$) atomic layer). In contrast, only ($\sqrt{3}\times\sqrt{3}$) structure can be identified on these films imaged at a sample bias of $-1.0\ \textrm{V}$.

To account for the bias and thickness dependent topography, STS spectra are taken on films of varying thickness.  As shown in Figs.~3(a) and (b), two pronounced filled states at $\sim-\textrm{0.2} \ e\textrm{V}$ and $\sim-\textrm{0.9} \ e\textrm{V}$ can be identified on all atomic layers. Meanwhile, as the layer thickness increases, both states shift slightly towards the Fermi level, but the energy difference between the two remains constant ($\sim\textrm{0.7} \ e\textrm{V}$). It is worth noting that this property of the ($\sqrt{3}\times\sqrt{3}$) reconstructed surface resembles that of the Si(111)-Ag($\sqrt{3}\times\sqrt{3}$)R$30^{\circ}$ surface, which exhibits a bonding state composed of the Ag 5$p$ orbits with the binding energy ranged between 0 and $-\textrm{0.3} \ e\textrm{V}$ with respect to the Fermi level (denoted as $s\textrm{1}$), and a state stemming mainly from the Ag 5$s$ orbits that are centered around $-\textrm{1} \ e\textrm{V}$ (denoted as $s\textrm{2}/s\textrm{3}$)$^{37-40}$. It is also known that charge donation to the Si(111)-Ag($\sqrt{3}\times\sqrt{3}$)R$30^{\circ}$ surface by the presence of excess Ag atoms (beyond what is needed to form the reconstructed surface, i.e., $\sim\textrm{1} \ \textrm{ML}$) induces a peak shift of the surface states away from the Fermi level, while the energy difference between $s\textrm{1}$ and $s\textrm{2}/s\textrm{3}$ states is kept constant at $\textrm{0.7} \ e\textrm{V}$$^{41-44}$. The striking similarities between these two surfaces suggest that the ($\sqrt{3}\times\sqrt{3}$) reconstructed structure achieved by depositing Si on the Ag(111) substrate should intrinsically be a Ag-Si alloy via diffusion of Ag atoms from the substrate to the top surface through the bulk-like Si film. Similar to the Si(111)-Ag($\sqrt{3}\times\sqrt{3}$)R$30^{\circ}$ surface, the energy shift of the surface states can be attributed to electron donation by excess Ag atoms. We expect this carrier doping effect to decay as the film thickness increases, resulting in a gradual shift of surface state peaks toward the Fermi level.

Next, we discuss the bias-dependent imaging associated with this distinct surface electronic structure. When the surface is imaged at the $-\textrm{1} \ \textrm{V}$ sample bias, the $s\textrm{2}/s\textrm{3}$ surface state dominates the tunneling process, leading to the observation of the prevalent filled state image of the ($\sqrt{3}\times\sqrt{3}$) structure (see the STM images in Fig. S4 which were taken on the sixth ($\sqrt{3}\times\sqrt{3}$) atomic layer). However, when the surface is imaged at $+\textrm{1.5} \ \textrm{V}$, where the density of states is mainly contributed from the bulk-like Si film$^{38,40}$, the majority of electrons from the tip will tunnel into the empty state of the bulk. Thus, the corresponding STM images are governed by the atomic structure of the film as well as the interference pattern between the top surface and the bottom interface, showing both the ($\sqrt{3}\times\sqrt{3}$) reconstruction and the ($\sqrt{21}\times\sqrt{21}$) superstructure, i.e., Moir\'{e} pattern, as presented in Figs.~2(a), (d), (g) and (j), especially in the thin film case. Intriguingly, at a sample bias of $+\textrm{0.5} \ \textrm{V}$, the Moir\'{e} pattern is only pronounced in the images taken on the first and second atomic layers (Figs.~2(b) and (e)), while on thicker films the ($\sqrt{3}\times\sqrt{3}$) periodicity dominates (Figs.~2(h) and (k)), presumably due to the enhanced $s\textrm{1}$ state and the associated electron tunneling to $s\textrm{1}$, resulting in the observation of the empty state image of the ($\sqrt{3}\times\sqrt{3}$) structure (see the STM images in Fig. S4 which were taken on the sixth ($\sqrt{3}\times\sqrt{3}$) atomic layer).

To confirm that the ($\sqrt{21}\times\sqrt{21}$) periodicity indeed originates from Moir\'{e} interference, we performed a calculation on the geometric structure and the surface diffraction pattern, assuming that the ($\sqrt{3}\times\sqrt{3}$)R$30^{\circ}$ adlayer is grown above the ($\sqrt{7}\times\sqrt{7}$)R$19.1^{\circ}$ interface structure$^{14}$. Note that these periodicities are all indexed with regard to the Si lattice. As shown in Fig.~S5, when the primary unit vectors of the two lattices are azimuthally rotated to each other by $\textrm{10.9}^{\circ}$, a ($\sqrt{21}\times\sqrt{21}$) superstructure is generated in both the surface electron diffraction pattern and the simulated geometric structure, suggesting a well-defined epitaxial relation between the two. Note that the FFT images can be directly compared with the surface diffraction pattern as they both probe the reciprocal space of the lattices. Here, the observation of the Moir\'{e} pattern indicates that the ($\sqrt{3}\times\sqrt{3}$) phase tends to grow on top of the ($\sqrt{7}\times\sqrt{7}$) interface structure instead of directly on the Ag(111) substrate. During this process, the nearby ($\sqrt{7}\times\sqrt{7}$) material is being consumed$^{31, 34}$. Finally, we note that the ($\sqrt{21}\times\sqrt{21}$) Moir\'{e} pattern fades away as the interface/substrate influence subsides, which is evidenced by the diminished ($\sqrt{21}\times\sqrt{21}$) signal in the thicker multilayered films shown in Fig.~2 and Fig.~S4.

As is well known, the s1 state of the analogous  Si(111)-Ag($\sqrt{3}\times\sqrt{3}$)R$30^{\circ}$ surface exhibits a parabolic dispersion crossing the Fermi level that is located within the projected bulk band gap, enabling the propagation of a surface two-dimensional electron gas (2DEG)$^{37-39}$. Although there have been reports on the energy-momentum dispersion relation of the ($\sqrt{3}\times\sqrt{3}$) phase of few-layer Si grown on Ag(111) by differential conductance (dI/dV) mapping of surface standing wave patterns, these measurements were performed over very limited energy ranges$^{25-28}$. Consequently, linear dispersion relations have been claimed and attributed to the existence of Dirac fermions. Here, we conduct similar measurements by dI/dV mapping, but over a wide energy range, on a ($\sqrt{3}\times\sqrt{3}$) reconstructed Si film with thickness larger than six atomic layers. Note that in this letter this is the only experiment carried out at the liquid helium temperature ($4.5\ \textrm{K}$), which is crucial for probing the scattering of low-energy surface electrons. Compared to the images of the surface taken at $77\ \textrm{K}$, triangular-shaped domains separated by domain boundaries are observed, suggesting that a structural phase transition has occurred$^{27,29}$. Nevertheless, these domain boundaries do not seem to act as scattering centers for surface electrons. Rather, scattering is dominated by the step edges, resulting in distinct standing wave patterns at different sample biases, as revealed in Fig.~4 (c)-(f). The layout of the participating step edges is illustrated in a zoomed-out STM image as shown in Fig.~S7(a). With the wave numbers of the surface electrons at different energies further determined from the FFT images of the corresponding dI/dV maps, we are able to derive a parabolic dispersion relation of the free-electron like surface state. The effective electron mass, as deduced from the dispersion curvature, is $m^*$= ($0.15\pm0.01$)$m_e$, where $m_e$ is the free electron mass. These results are consistent with those obtained on the  Si(111)-Ag($\sqrt{3}\times\sqrt{3}$)R$30^{\circ}$ surface$^{45}$. Thus, we can conclude that the ($\sqrt{3}\times\sqrt{3}$) phase of few-layer Si grown on Ag(111) is not a Dirac fermion system as claimed by Chen et al.$^{25-28}$, instead it shows the 2DEG properties identical to Si(111)-Ag-($\sqrt{3}\times\sqrt{3}$)R$30^{\circ}$. It is worth noting that such a parabolic dispersion was previously observed, although it was originally attributed to the Ag(111) surface state$^{29}$.

Built on the profound understanding of the surface electronic structure and the origin of the surface 2DEG, we next explore the evolution of the s1 state, specifically its magnitude and line width, as a function of layer thickness. As shown in Fig.~3(b), the s1 state is suppressed on the first ($\sqrt{3}\times\sqrt{3}$) atomic layer, while on layers further away from the interface the corresponding STS peak becomes more pronounced and sharper. To comprehend the trend, we perform a series of dI/dV mapping on ($\sqrt{3}\times\sqrt{3}$) reconstructed films of varying thickness on Ag(111). The sample bias is chosen between $+0.3\ \textrm{V}$ and $+0.9\ \textrm{V}$, allowing standing waves to be pronounced at $77\ \textrm{K}$.

Figs.~5(a) and (b) show the surface topography and corresponding differential conductance map, respectively, taken along the boundary between the bare Ag(111) substrate and the first ($\sqrt{3}\times\sqrt{3}$) atomic layer (yellow square in Fig.~1(a)). Although standing wave oscillations through the Shockley surface state on Ag(111) are readily observed$^{46}$, the wave oscillations are too weak to be distinguished on the ($\sqrt{3}\times\sqrt{3}$) structure. This is further corroborated by the line profile, as shown in Fig.~5(e), taken along the $\textrm{A}-\textrm{B}$ section illustrated in Fig.~5(b). In contrast, standing wave oscillations start to emerge on the second ($\sqrt{3}\times\sqrt{3}$) atomic layer, accompanied by an increased wave magnitude and wave decay length with layer number. As illustrated in the differential conductance map in Fig.~5(d) (the corresponding topography shown in Fig.~5(c)) and the line profile across the boundary of the second and third ($\sqrt{3}\times\sqrt{3}$) atomic layers (red lines in Figs.~5(d) and (e)), only three weak standing wave oscillation peaks are observed on the second layer, while it shows four strong oscillations on the third layer. The zoomed-out STM image of the same area is presented in Fig.~S9(a), where the layer numbers can be precisely identified. A similar example of the strengthened standing wave pattern on upper layers is depicted in Figs.~S8(d) and (e), and as well in Figs.~S9(c) and (d). Note that the potential difference in the strength of the scattering barrier at the vacuum-film vs. the film-film edge and how it contributes to the standing wave oscillation have been thoroughly evaluated, with the details provided in Fig.~S10.

It is known that the phase coherence length of a 2DEG, $\textrm{L}_{\phi}$, resulting from the free-electron-like surface state, is proportional to the decay length of the surface standing wave and inversely proportional to the line width of the STS peak$^{47-50}$. The similarity in the thickness-dependent trend between the evolution of the $s\textrm{1}$ state and the standing wave pattern further confirms that the 2DEG observed on the ($\sqrt{3}\times\sqrt{3}$) phase originates from the $s\textrm{1}$ surface state. Moreover, the thickness-dependent $\textrm{L}_{\phi}$ strongly reflects a substrate influence on the properties of the 2DEG. We speculate that the inelastic inter-band electron-electron ($e$-$e$) scattering resulting from the overlap between the $s\textrm{1}$ surface state, interface state of the ($\sqrt{7}\times\sqrt{7}$) superstructure, and the bulk state of the Ag substrate is the main driving force for the decayed/diminished standing wave observed on the ($\sqrt{3}\times\sqrt{3}$) structures closely atop the Ag(111) substrate$^{48,51,52}$. A similar scattering mechanism has led to the decay of standing waves on noble metal surfaces when the surface state band approaches the bulk band edge$^{48,52}$.

The strong substrate effect is also reflected in the electronic structure of the ($\sqrt{7}\times\sqrt{7}$) pattern. Fig.~6(a) shows the STS spectra taken on the bare Ag(111) surface and on the ($\sqrt{7}\times\sqrt{7}$) surface. The Shockley surface state rising at $\sim-\textrm{60} \ \textrm{mV}$ accounts for the 2DEG observed on the Ag(111) surface, as illustrated in Fig.~5(b), Figs.~6(b) and (c), and Figs.~S8 (a)-(c). However, this state is smeared by the ($\sqrt{7}\times\sqrt{7}$) structure, likely owing to the strong hybridization between the adlayer and the substrate surface$^{21-24}$, leading to the formation of a new interface state at around $\textrm{0.2} \ e\textrm{V}$ above the Fermi level. The differential conductance mapping images shown in Figs.~6(b) and (c) further illustrate the absence of standing waves underneath the ($\sqrt{7}\times\sqrt{7}$) structure, suggesting that the free-electron-like Ag(111) surface state has been eliminated$^{53}$. Thus, in contrast to the earlier report$^{29}$, our observation implies that the 2DEG observed on the ($\sqrt{3}\times\sqrt{3}$) phase grown on top of the ($\sqrt{7}\times\sqrt{7}$) structure cannot originate from the Ag(111) surface state.

\section*{Conclusion}

\noindent We have investigated the influence of interfacial coupling on the electronic structure of the few-layer Si grown on the Ag(111) surface. These films display the bulk-like $sp^3$ hybridization, while the ($\sqrt{3}\times\sqrt{3}$) reconstructed surface interferes with the ($\sqrt{7}\times\sqrt{7}$) interfacial structure resulting in a ($\sqrt{21}\times\sqrt{21}$) Moir\'{e} pattern. Combining STS and differential conductance mapping, we show that the free-electron-like surface state of the ($\sqrt{3}\times\sqrt{3}$) structure gradually diminishes, associated with a decrease in electron phase coherence length when approaching the interface. These features are presumably attributable to the inelastic inter-band electron-electron scattering originating from the overlap between the surface state, interface state and the bulk state of the substrate. We further demonstrate that the intrinsic electronic structure of the as grown ($\sqrt{3}\times\sqrt{3}$) phase is identical to that of the ($\sqrt{3}\times\sqrt{3}$)R$30^{\circ}$ reconstructed Ag on Si(111), both of which exhibit the parabolic energy-momentum dispersion relation with comparable electron effective masses. These findings highlight the essential role of interfacial coupling on the properties of two-dimensional Si structures grown on supporting substrates, which should be thoroughly scrutinized in pursuit of silicene.

\section*{Methods}

\noindent The experiments were carried out in an Omicron NanoTechnology GmbH low-temperature scanning tunneling microscope (LT-STM) equipped with a separate sample preparation system. The base pressures of the two chambers were both maintained below $\textrm{1}\times\textrm{10}^{-\textrm{10}} \ \textrm{mbar}$. The silver (Ag) substrate was cleaned in the preparation system by argon ion sputtering ($\textrm{1} \ \textrm{k}e\textrm{V}/\textrm{25} \ \textrm{A}$) for 30 minutes followed by thermal annealing at $\sim\textrm{500}^\circ$C for several cycles. Si was then evaporated from a custom-built evaporator ($\sim\textrm{1000}^\circ$C) onto the preheated Ag substrate ($\sim\textrm{320}^\circ$C) with a growth rate of $\sim\textrm{0.05} \ \textrm{ML}/\textrm{min}$. After deposition, the sample was \textit{in situ} transferred to the LT-STM chamber and cooled to $77\ \textrm{K}$ or $4.5\ \textrm{K}$ for STM/STS measurements. STS and differential conductance mapping were obtained by applying a small modulation signal (18 mV) at the frequency of 1.2 kHz to the tip-sample junction and detecting the corresponding ac tunneling current signal with a lock-in amplifier. Spectra on Ag(111) were taken periodically as a reference to confirm tip consistency.

\section*{References}

\begin{enumerate}

\item Takeda, K. \& Shiraishi, K. Theoretical possibility of stage corrugation in Si and Ge analogs of graphite. \textit{Phys. Rev. B} \textbf{50}, 14916 (1994).

\item Cahangirov, S., Topsakal, M., Akt\"{u}rk, E., \c{S}ahin, H. \& Ciraci, S. Two- and one-dimensional honeycomb structures of silicon and germanium. \textit{Phys. Rev. Lett.} \textbf{102}, 236804 (2009).

\item Guzm\'{a}n-Verri, G. G. \& Voon, L. Y. Electronic structure of silicon-based nanostructures. \textit{Phys. Rev. B} \textbf{76}, 075131 (2007).

\item Ni, Z. \textit{et al}. Tunable bandgap in silicene and germanene. \textit{Nano Lett.} \textbf{12}, 113 (2012).

\item Ezawa, M. Valley-polarized metals and quantum anomalous Hall effect in silicene. \textit{Phys. Rev. Lett.} \textbf{109}, 055502 (2012).

\item Liu, C.-C., Feng, W. \& Yao, Y. Quantum spin Hall effect in silicene and two-dimensional germanium. \textit{Phys. Rev. Lett.} \textbf{107}, 076802 (2011).

\item Liu, C.-C., Jiang, H. \& Yao, Y. Low-energy effective Hamiltonian involving spin-orbit coupling in silicene and two-dimensional germanium and tin. \textit{Phys. Rev. B} \textbf{84}, 195430 (2011).

\item Vogt, P. \textit{et al}. Silicene: Compelling experimental evidence for graphene-like two-dimensional silicon. \textit{Phys. Rev. Lett.} \textbf{108}, 155501 (2012).

\item Feng, B. \textit{et al}. Evidence of silicene in honeycomb structures of silicon on Ag(111). \textit{Nano Lett.} \textbf{12}, 3507 (2012).

\item Liu, Z. L. \textit{et al}. The fate of the ($2\sqrt{3}\times2\sqrt{3}$)R$30^{\circ}$ silicene phase on Ag(111). \textit{APL Mat.} \textbf{2}, 092513 (2014).

\item Salomon, E. \& Kahn, A. One-dimensional organic nanostructures: A novel approach based on the selective adsorption of organic molecules on silicon nanowires. \textit{Surf. Sci.} \textbf{602}, L79 (2008).

\item D\'{a}vila, M. E. \textit{et al}. Comparative structural and electronic studies of hydrogen interaction with isolated versus ordered silicon nanoribbons grown on Ag(110). \textit{Nanotechnology} \textbf{23}, 385703 (2012).

\item Kara, A. \textit{et al}. A review on silicene--New candidate for electronics. \textit{Surf. Sci. Rep.} \textbf{67}, 1 (2012).

\item Jamgotchian, H. \textit{et al}. A comprehensive study of the ($2\sqrt{3}\times2\sqrt{3}$)R$30^{\circ}$ structure of silicene on Ag(111). \textit{e-print arXiv:1412.4902} (2014).

\item Resta A. \textit{et al.} Atomic structures of silicene layers grown on Ag(111): Scanning tunneling microscopy and noncontact atomic force microscopy observations. \textit{Sci. Rep.} \textbf{3}, 2399 (2013).

\item Arafune, R.  \textit{et al}. Structural transition of silicene on Ag(111).  \textit{Surf. Sci.} \textbf{608}, 297 (2013).

\item De Padova P.  \textit{et al}. 24 h stability of thick multilayer silicene in air.  \textit{2D Mater.} \textbf{1}, 021003 (2014).
\item Sone, J. \textit{et al}. Epitaxial growth of silicene on ultra-thin Ag(111) films.  \textit{New J.  Phys.} \textbf{16}, 095004 (2014).

\item Fleurence, A. \textit{et al}. Experimental evidence for epitaxial silicene on diboride thin films. \textit{Phys. Rev. Lett.} \textbf{108}, 245501 (2012).

\item Meng, L. \textit{et al}. Buckled silicene formation on Ir(111). \textit{Nano Lett.} \textbf{13}, 685 (2013).

\item Gou, Z. X., Furuya, S., Iwata, J. \& Oshiyama, A. Absence and presence of Dirac electrons in silicene on substrates. \textit{Phys. Rev. B} \textbf{87}, 235435 (2013).

\item Lin, C. L. \textit{et al}. Substrate-induced symmetry breaking in silicene. \textit{Phys. Rev. Lett.} \textbf{110}, 076803 (2013).

\item Cahangirov, S. \textit{et al}. Electronic structure of silicene on Ag(111): Strong hybridization effects. \textit{Phys. Rev. B} \textbf{88}, 035432 (2013).

\item Tsoutsou, D., Xenogiannopoulou, E., Golias, E., Tsipas, P. \& Dimoulas, A. Evidence for hybrid surface metallic band in ($4\times4$) silicene on Ag(111). \textit{Appl. Phys. Lett.} \textbf{103}, 231604 (2013).

\item Chen, L.  \textit{et al}. Evidence for Dirac fermions in a honeycomb lattice based on silicon. \textit{Phys. Rev. Lett.} \textbf{109}, 056804 (2012).

\item Feng, B. \textit{et al}. Observation of Dirac cone warping and chirality effects in silicene. \textit{ACS Nano} \textbf{7}, 9049 (2013).

\item Chen, L. \textit{et al}. Spontaneous symmetry breaking and dynamic phase transition in monolayer silicene. \textit{Phys. Rev. Lett.} \textbf{110}, 085504 (2013).

\item Chen, J. \textit{et al}. Persistent Dirac fermion state on bulk-like Si(111) surface.  \textit{e-print arXiv:1405.7534} (2014).

\item Arafune, R., Lin, C. L., Nagao, R., Kawai, M. \& Takagi, N. Comment on ``Evidence for Dirac fermions in a honeycomb lattice based on silicon''. \textit{Phys. Rev. Lett.} \textbf{110}, 229701 (2013).

\item Padoca, P. \textit{et al}. Evidence of Dirac fermions in multilayer silicene. \textit{Appl. Phys. Lett.} \textbf{102}, 163106 (2012).

\item Mannix, A. J., Kiraly, B., Fisher, B. L., Hersam, M. C. \& Guisinger, N. P. Silicon growth at the two-dimensional limit on Ag(111). \textit{ACS Nano} \textbf{8}, 7538 (2014).

\item  Yamagami, T., Sone, J., Nakatsuji, K. \& Hirayama, H. Surfactant role of Ag atoms in the growth of Si layers on Si(111) ($\sqrt{3}\times\sqrt{3}$)-Ag substrates. \textit{Appl. Phys. Lett.} \textbf{105}, 151603 (2014).

\item  Shirai, T. \textit{et al}. Structure determination of multilayer silicene grown on Ag(111) films by electron diffraction: Evidence for Ag segregation at the surface. \textit{Phys. Rev. B} \textbf{89}, 241403(R) (2014).

\item Acun, A., Poelsema, B., Zandvliet, H. J. W. \& van Gastel, R. The instability of silicene on Ag(111). \textit{Appl. Phys. Lett.} \textbf{103}, 263119 (2013).

\item Jamgotchian, H. \textit{et al}. Growth of silicene layers on Ag(111): Unexpected effect of the substrate temperature. \textit{J. Phys.: Condens. Matter} \textbf{24}, 172001 (2012).

\item  Rahman, M. S.,  Nakagawa, T. \& Mizuno, S. Growth of Si on Ag(111) and determination of large commensurate unit cell of high-temperature phase. \textit{Jpn. J. Appl.  Phys.} \textbf{54}, 015502 (2015).

\item Aizawa, H. \& Tsukada, M. First-principles study of Ag adatoms on the Si(111)-Ag($\sqrt{3}\times\sqrt{3}$)R$30^{\circ}$ surface. \textit{Phys. Rev. B} \textbf{59}, 10923 (1999).

\item Hasegawa, S. \textit{et al}. Surface state bands on silicon-Si(111)-Ag surface superstructure. \textit{Jpn. J. Appl. Phys.} \textbf{39}, 3815 (2000).

\item Johansson, L. S. O., Landemark, E., Karlsson, C. J. \& Uhrberg, R. I. G. Fermi-level pinning and surface-state band structure of the Si(111)-Ag($\sqrt{3}\times\sqrt{3}$)R$30^{\circ}$ surface. \textit{Phys. Rev. Lett.} \textbf{63}, 2092 (1989).

\item Wan, K., Lin, X. F. \& Nogami, J. Reexamination of the Ag/Si(111)-$\sqrt{3}\times\sqrt{3}$ surface by scanning tunneling microscopy. \textit{Phys. Rev. B} \textbf{45}, 9509 (1992).

\item Nakajima, Y., Takeda, S., Nagao, T., Hasegawa, S. \& Tong, X. Surface electrical conduction due to carrier doping into a surface-state band on Si(111)-($\sqrt{3}\times \sqrt{3}$)-Ag. \textit{Phys. Rev. B} \textbf{56}, 6782 (1997).

\item Tong, X., Jiang, C. S. \& Hasegawa, S. Electronic structure of the Si(111)-($\sqrt{21}\times\sqrt{21}$)-(Ag$+$Au) surface. \textit{Phys. Rev. B} \textbf{57}, 9015 (1998).

\item Uhrberg, R. I. G., Zhang, H. M., Balasubramanian, T., Landemark, E. \& Yeom, H. W. Photoelectron spectroscopy study of Ag/Si(111)-$\sqrt{3}\times\sqrt{3}$ and the effect of additional Ag adatoms. \textit{Phys. Rev. B} \textbf{65}, 081305(R) (2002).

\item Ono, M., Nishigata, Y.,  Nishio, T.,  Eguchi, T. \& Hasegawa, Y. Electrostatic potential screened by a two-dimensional electron system: A real-space observation by scanning-tunneling spectroscopy. \textit{Phys. Rev. Lett.} \textbf{96}, 016801 (2006).

\item  Hirahara, T.,  Matsuda, I.,  Ueno, M. \& Hasegawa, S. The effective mass of a free-electron-like surface state of the Ag/Si(111)-$\sqrt{3}\times\sqrt{3}$  surface investigated by photoemission and scanning tunneling spectroscopies. \textit{Surf. Sci.} \textbf{563}, 191 (2004).

\item Pennec, Y. \textit{et al}. Supramolecular gratings for tunable confinement of electrons on metal surfaces. \textit{Nat. Nanotechnol.} \textbf{2}, 99 (2007).

\item B\"{u}rgi, L., Jeandupeux, O., Brune, H. \& Kern, K. Probing hot-electron dynamics at surfaces with a cold scanning tunneling microscope. \textit{Phys. Rev. Lett.} \textbf{82}, 4516 (1999).

\item Vitali, L. \textit{et al}. Inter- and intraband inelastic scattering of hot surface state electrons at the Ag(111) surface. \textit{Surf. Sci.} \textbf{523}, 47 (2003). 

\item Fukumoto, H., Aoki, Y. \& Hirayama, H. Decay of Shockley surface state by randomly adsorbed Bi atoms at Ag(111) surfaces. \textit{Phys. Rev. B} \textbf{86}, 165311 (2012).

\item Li, J., Schneider, W.-D., Berndt, R., Bryant, O. R. \& Crampin, S. Surface-state lifetime measured by scanning tunneling spectroscopy. \textit{Phys. Rev. Lett.} \textbf{81}, 4464 (1998).

\item Silkin, V. M., Balassis, A., Leonardo, A., Chulkov, E. V. \& Echenique, P. M. Dynamic screening and electron dynamics in non-homogeneous metal systems. \textit{Appl. Phys. A} \textbf{92}, 453 (2008).

\item \"{U}nal, A. A. \textit{et al}. Hybridization between the unoccupied Shockley surface state and bulk electronic states on Cu(111). \textit{Phys. Rev. B} \textbf{84}, 073107 (2011).

\item Mart\'{i}nez-Galera, A. J., Brihuega, I. \& G\'{o}mez-Rodr\'{i}guez, J. M. Ethylene irradiation: A new route to grow graphene on low reactivity metals. \textit{Nano Lett.} \textbf{11}, 3576 (2011).

\end{enumerate}

\section*{Acknowledgments}

\noindent Research supported by the National Science Foundation (NSF) under Award \#DMR-1410417 (synthesis of samples and scanning tunneling microscopy/spectroscopy studies), and by Department of Energy (DOE) Office of Science Early Career Research Program through the Office of Basic Energy Sciences, Award \#DE-SC0006400 (calculations on surface diffraction pattern and geometric structure).

\section*{Author contributions}

\noindent P.P.Z. and J.F. conceived the project; J.F. performed the experiments, and analyzed the data with P.P.Z.; S.R.W. carried out calculations on surface diffraction pattern and geometric structure; P.P.Z., J.F., and S.R.W. wrote the manuscript and discussed the results.

\section*{Additional information}

\noindent \textbf{Supplemental information} accompanies this paper at\newline \href{}{http://www.nature.com/scientificreports}\newline 
\noindent \textbf{Competing financial interests:} The authors declare no competing financial interests.\newline
\noindent \textbf{Further information:} Correspondence and requests for materials should be addressed to P.P.Z.\newline
\noindent \textbf{How to cite this article:} Feng, J. \textit{ et al.} Interfacial Coupling and Electronic Structure of Two-Dimensional Silicon Grown on the Ag(111) Surface at High Temperature. \textit{ Sci. Rep.}  \textbf{5}, 10310; doi: 10.1038/srep10310 (2015).

\newpage

\begin{figure*}[ht!]
\center
\includegraphics[width=1.0\textwidth,angle=0]{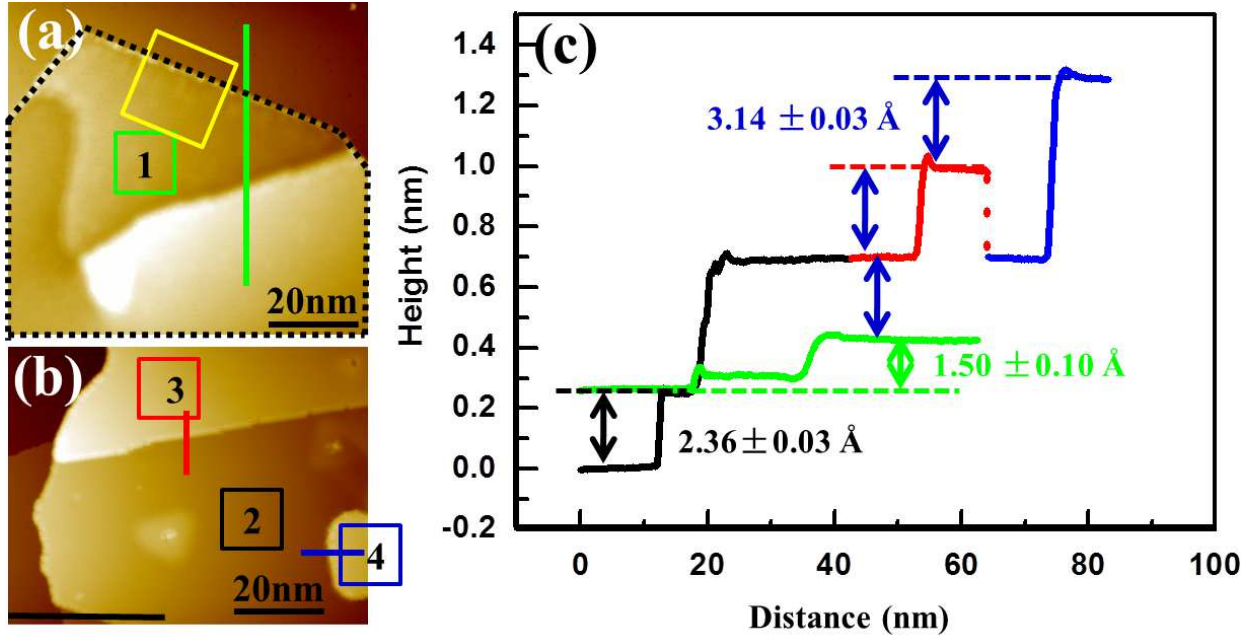}
\caption{\textbf{STM topography images and line profiles obtained at $77\ \textrm{K}$ of few-layer Si with the ($\sqrt{3}\times \sqrt{3}$) phase on Ag(111).} (a) STM image shows the first atomic layer and (b) multilayered Si with the ($\sqrt{3}\times\sqrt{3}$) reconstructed surface on the Ag(111) substrate ($V_{s}=+\textrm{1.5} \ \textrm{V}$; $I_{t}=\textrm{50} \ \textrm{pA}$). The corresponding layer numbers are labeled in the boxed regions of (a) and (b), which are also used as reference locations for zoomed-in STM topography images discussed later. Note that the area outlined by the black dotted line in (a) presents a continuous film. (c) Apparent height line profiles taken along the green, red, blue, and black marks denoted in (a) and (b) with the appropriate corresponding color. The line profiles show the apparent interlayer spacing of Si structures (blue arrows), the apparent height difference between the first ($\sqrt{3}\times\sqrt{3}$) atomic layer and the underlying Ag surface (the green arrow), and the apparent out-of-plane spacing of the Ag(111) substrate (the black arrow). The interlayer spacing of Si structures measured by STM matches with the d-spacing of bulk Si(111).}
\label{fig:1}
\end{figure*}

\newpage

\begin{figure*}[ht!]
\center
\includegraphics[width=0.7\textwidth,angle=0]{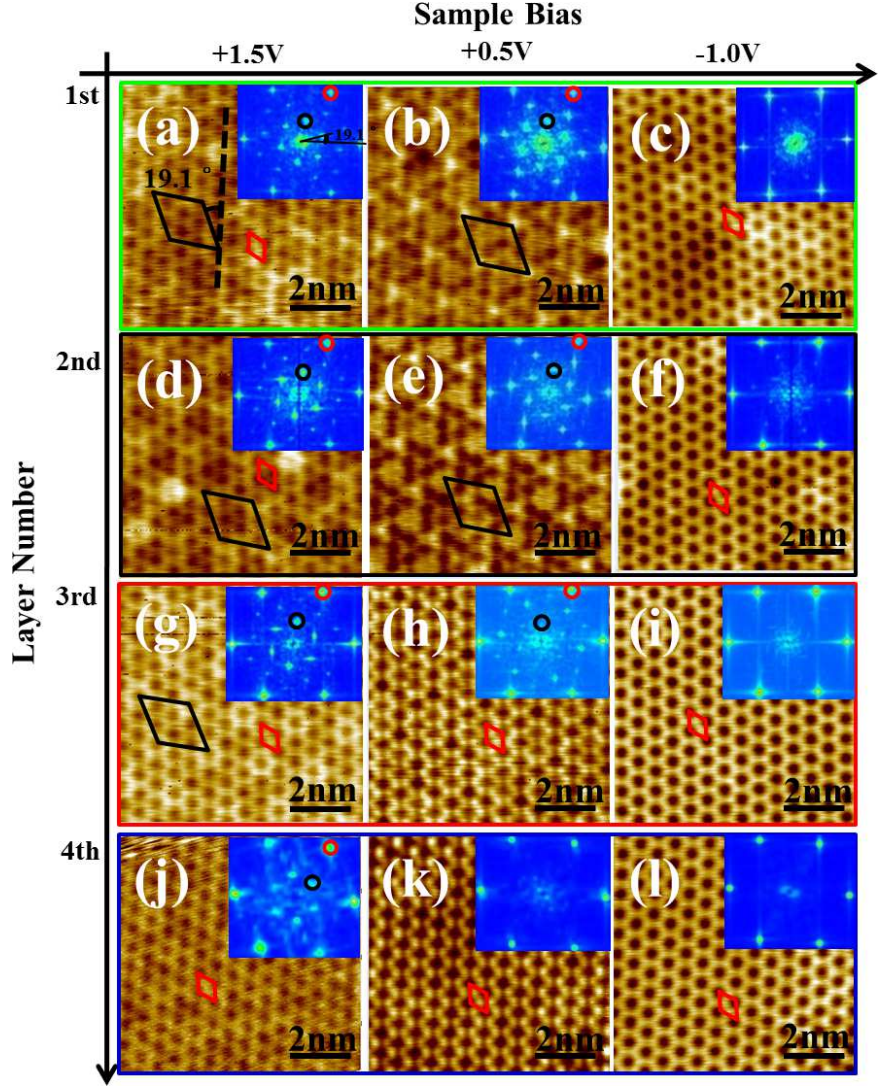}
\caption{\textbf{Bias- and thickness- dependent surface topographies of the ($\sqrt{3}\times\sqrt{3}$) phase obtained at $77\ \textrm{K}$.} (a)-(l) A series of STM topography images ($V_{s}=+\textrm{1.5} \ \textrm{V}$, $+\textrm{0.5} \ \textrm{V}$, $-\textrm{1.0} \ \textrm{V}$; $I_{t}=\textrm{50} \ \textrm{pA}$) of the topmost atomic layer of the ($\sqrt{3}\times\sqrt{3}$) phase on the Ag(111) substrate. The STM images are arranged such that all images within a given column have the same sample bias corresponding to the value given at the top of the figure, while images in each row corresponds to the same atomic layer labeled on the left side of the figure. The green, red, blue, and black outline of each row corresponds to the boxed regions in Fig.~1. The insets in each image represent the corresponding FFT of the STM image. The ($\sqrt{3}\times\sqrt{3}$) reconstruction (red) and the ($\sqrt{21}\times\sqrt{21}$) superstructure (black) are observed and rotated by $\textrm{19.1}^{\circ}$ with respect to each other. The corresponding unit cell and FFT spot are color labeled respectively.}
\label{fig:2} 
\end{figure*}

\newpage

\begin{figure*}[ht!]
\center
\includegraphics[width=1.0\textwidth,angle=0]{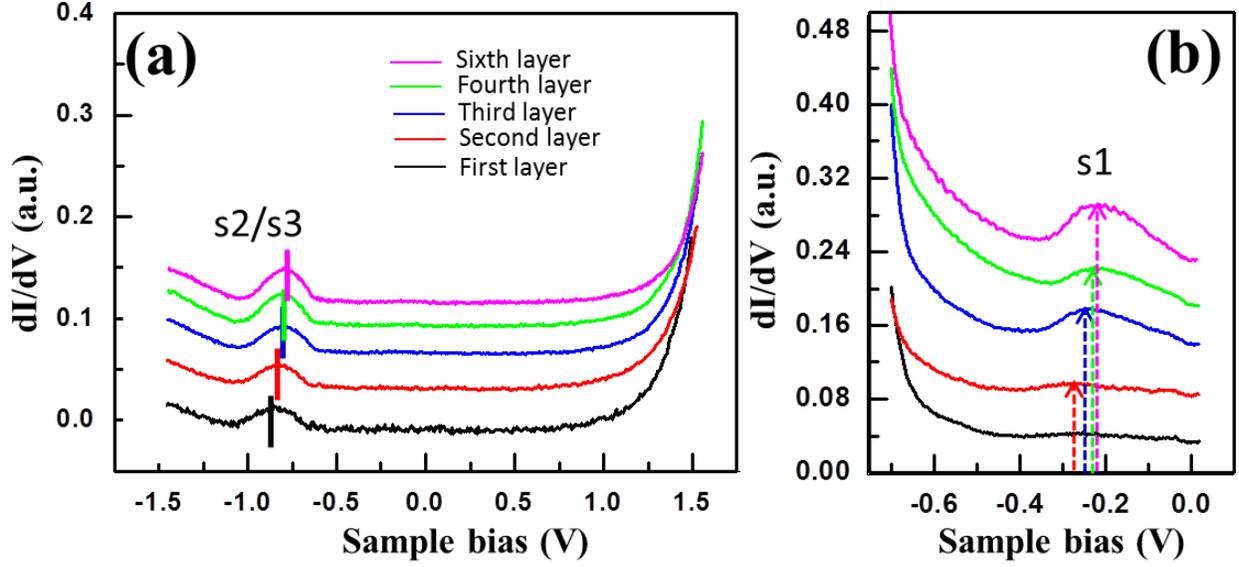}
\caption{\textbf{Evolution of the surface electronic structure of the ($\sqrt{3}\times \sqrt{3}$) phase with film thickness.} A series of STS $dI/dV$ spectra ($I_{t}=\textrm{300} \ \textrm{pA}$) obtained on different atomic layers of the ($\sqrt{3}\times\sqrt{3}$) phase on the Ag(111) substrate at $77\ \textrm{K}$. The sample bias sweep of the spectra shown in (a) ranges from $V_{s}=-\textrm{1.5} \ \textrm{V}$ to $+\textrm{1.5} \ \textrm{V}$, while the spectra in (b) ranges from $V_{s}=-\textrm{0.7} \ \textrm{V}$ to $\textrm{0} \ \textrm{V}$. The peaks observed at $-\textrm{0.9} \ \textrm{V}$ and $-\textrm{0.2} \ \textrm{V}$ correspond to the $s\textrm{2}/s\textrm{3}$ and $s\textrm{1}$ states, respectively. The curves are offset vertically to show the gradual change of the surface states with respect to the layer number.}
\label{fig:3}
\end{figure*}

\newpage

\begin{figure*}[ht!]
\center
\includegraphics[width=1.0\textwidth,angle=0]{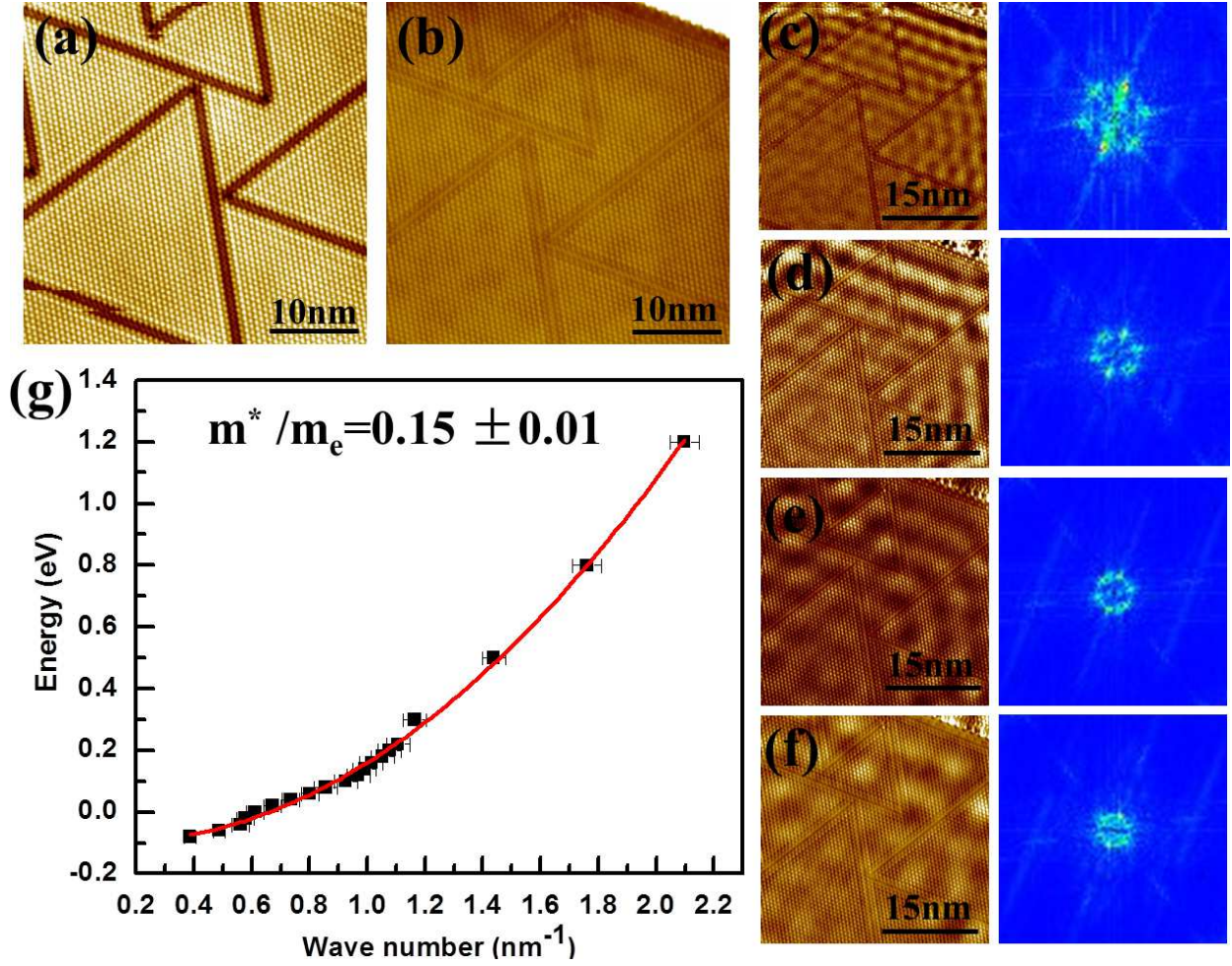}
\caption{\textbf{ STM topography images and differential conductance (dI/dV) mapping images of the ($\sqrt{3}\times\sqrt{3}$)  phase, obtained at $4.5\ \textrm{K}$, and the energy-momentum dispersion relation derived from the dI/dV mapping.}  (a) and (b) are STM topography images of a multilayered Si film ($>$ 6 atomic layers) with  the ($\sqrt{3}\times\sqrt{3}$) reconstructed surface imaged at ($V_{s}=+\textrm{1.2} \ \textrm{V}$; $I_{t}=\textrm{300} \ \textrm{pA}$) and ($V_{s}=\textrm{-0.04} \ \textrm{V}$; $I_{t}=\textrm{300} \ \textrm{pA}$) respectively. (c)-(f) dI/dV  maps and their corresponding FFT images, collected on the same area as shown in (a) and (b). $V_S$ is set at (c) $+0.3\ \textrm{V}$, (d) $+ 0.1\ \textrm{V}$, (e) $+0.04\ \textrm{V}$ and, (f) $-0.04\ \textrm{V}$, respectively, and $I_t$ at 300pA for all maps. (g) shows energy-momentum dispersion relation determined from the dI/dV mapping. Black squares with error bars represent experimental data and the red curve is the parabolic fitting}
\label{fig:4}
\end{figure*}

\newpage

\begin{figure*}[ht!]
\center
\includegraphics[width=1.0\textwidth,angle=0]{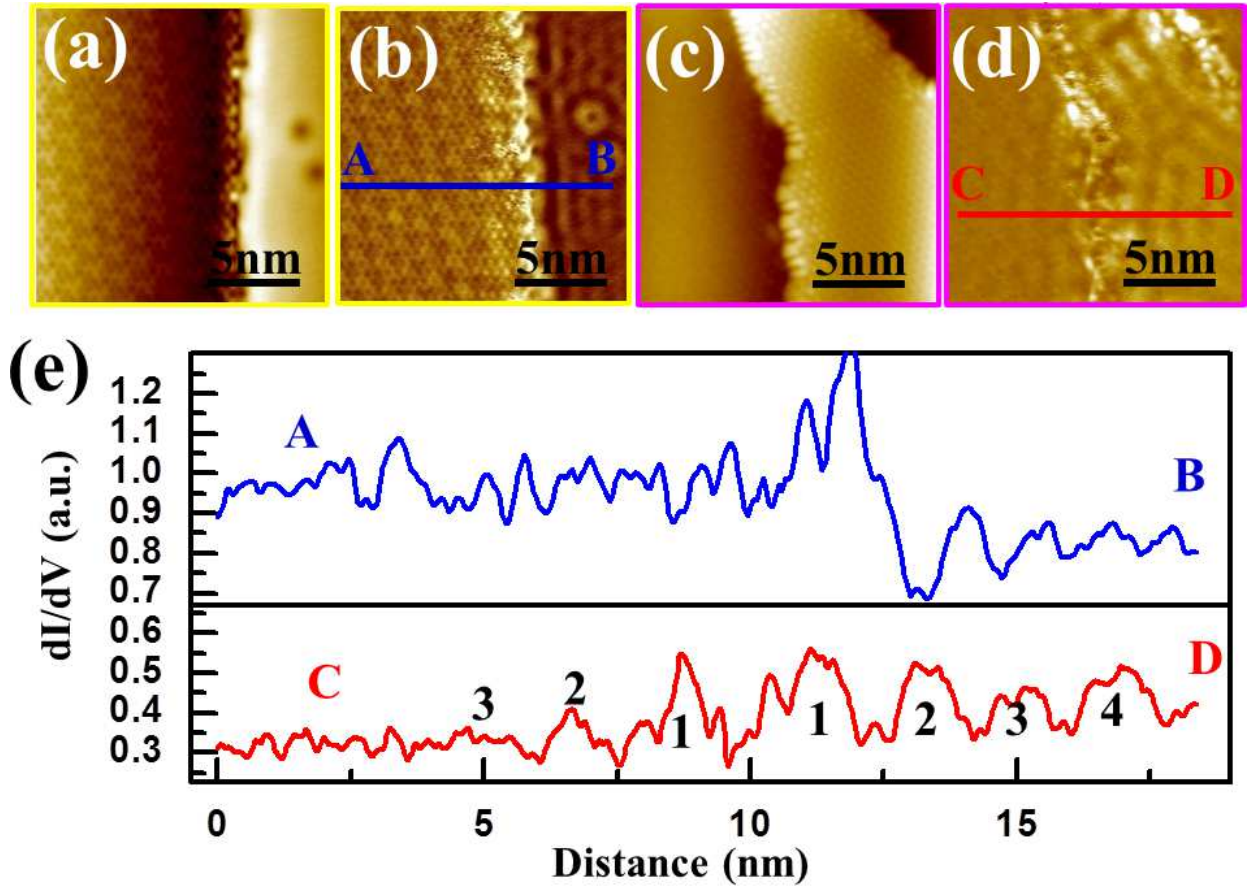}
\caption{\textbf{Differential conductance mapping images revealing the existence of two-dimensional electron gas on upper layers of the ($\sqrt{3}\times \sqrt{3}$) phase obtained at $77\ \textrm{K}$.} (a) and (b) are simultaneously obtained STM topography and differential conductance mapping images ($V_{s}=+\textrm{0.5} \ \textrm{V}$; $I_{t}=\textrm{300} \ \textrm{pA}$) taken on the first ($\sqrt{3}\times\sqrt{3}$) atomic layer on the Ag(111) substrate. The yellow outline around (a) and (b) highlights the region in Fig.~1 where the images were taken from. The standing wave oscillations on Ag(111) can be clearly identified in (b), however, the wave oscillations are too weak to be distinguished on the first ($\sqrt{3}\times\sqrt{3}$) layer as indicated by the line profile in (e) (blue line). (c) and (d) are simultaneously obtained STM topography and differential conductance images ($V_{s}=+\textrm{0.5} \ \textrm{V}$; $I_{t}=\textrm{300} \ \textrm{pA}$) taken on the second and third ($\sqrt{3}\times\sqrt{3}$) atomic layers on the Ag(111) substrate. The pink outline around (c) and (d) highlights the region in Fig. S9 where the images were taken from. Standing wave oscillations can be observed on both the second and third ($\sqrt{3}\times\sqrt{3}$) atomic layers, but they are stronger on the third layer, as indicated by the line profile in (e) (red line).}
\label{fig:5}
\end{figure*}

\newpage

\begin{figure*}[ht!]
\center
\includegraphics[width=1.0\textwidth,angle=0]{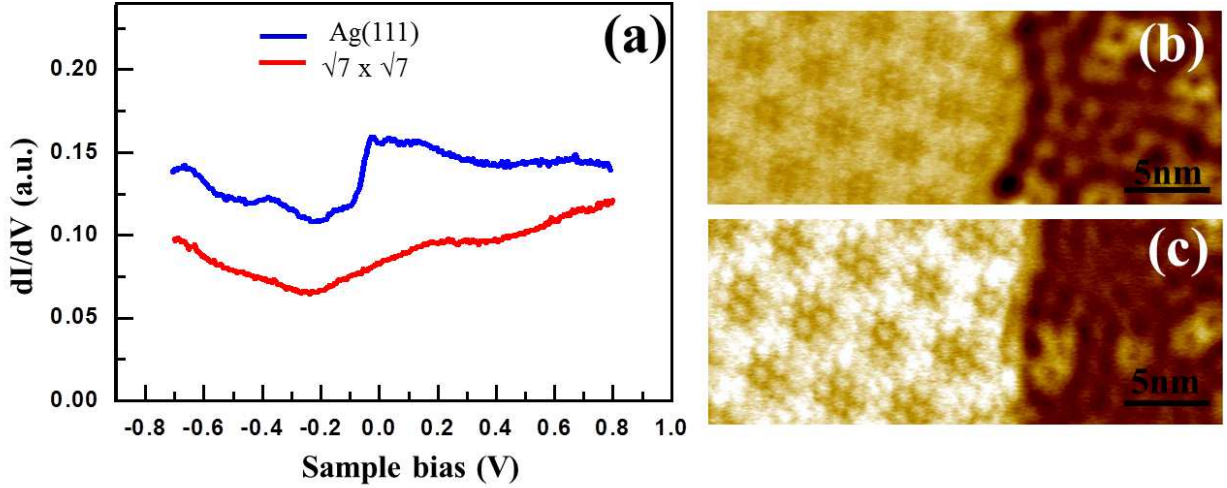}
\caption{\textbf{Electronic structure and differential conductance map on the monolayer ($\sqrt{7}\times\sqrt{7}$) superstructure obtained at $77\ \textrm{K}$.} (a) STS $dI/dV$ spectra (sample bias sweep from $V_{s}=-\textrm{0.7} \ \textrm{V}$ to $+\textrm{0.8} \ \textrm{V}$; $I_{t}=\textrm{300} \ \textrm{pA}$) obtained on the bare Ag(111) surface (blue) and on a single-layer of Si displaying the ($\sqrt{7}\times\sqrt{7}$) superstructure (red). The two curves are offset vertically for clarity. The sharp increase in the DOS of the bare Ag(111) surface at around $-\textrm{60} \ \textrm{mV}$ is attributed to the Shockley surface state, which is smeared out in the spectra taken on the ($\sqrt{7}\times\sqrt{7}$) superstructure, leading to a new interface state near $+\textrm{0.2} \ e\textrm{V}$ above the Fermi level. The differential conductance mapping images ($V_{s}=+\textrm{0.3} \ \textrm{V}$ in (b) and $+\textrm{0.5} \ \textrm{V}$ (c); $I_{t}=\textrm{300} \ \textrm{pA}$) show standing wave oscillations on the bare Ag(111) surface and the absence of wave oscillations underneath the ($\sqrt{7}\times\sqrt{7}$) superstructure.}
\label{fig:6}
\end{figure*}

\end{document}